\documentclass[12pt]{article}
\usepackage{amssymb,amsfonts}
\usepackage{graphics,psboxit,amsmath}
\usepackage{subfigure}
\usepackage{graphicx}
\usepackage{verbatim}
\usepackage{color}
\usepackage{hyperref}
\hypersetup{colorlinks}

\definecolor{darkred}{rgb}{1,0,0}
\definecolor{darkgreen}{rgb}{0,0.5,0}
\definecolor{darkblue}{rgb}{0,0,1}
\definecolor{orange}{rgb}{1,0.5,0}
\definecolor{green}{rgb}{0,1,0}
\definecolor{purple}{rgb}{.5,0,1}

\hypersetup{ colorlinks,
linkcolor=darkred,
filecolor=darkgreen,
urlcolor=darkblue,
citecolor=darkblue,
linktocpage=true }

\definecolor{markcolor}{rgb}{.25,0,1}

\definecolor{markcolor2}{rgb}{1,0,0}

\definecolor{markcolor3}{rgb}{0,1,0}

\def\hybrid{\topmargin -10pt    \oddsidemargin 0pt %%%%%%%%%%%%%% Archive-30pt
        \headheight 0pt \headsep 0pt
        \textwidth 16.5cm      % A4 paper
        \textheight 23cm       % A4 paper
        \marginparwidth .875in
        \parskip 5pt plus 1pt   \jot = 1.5ex}

\hybrid

\catcode`\@=11

\def\marginnote#1{}
\newcount\hour
\newcount\minute
\newtoks\amorpm
\hour=\time\divide\hour by60 \minute=\time{\multiply\hour by60
\global\advance\minute by-\hour}
\edef\standardtime{{\ifnum\hour<12 \global\amorpm={am}%
        \else\global\amorpm={pm}\advance\hour by-12 \fi
        \ifnum\hour=0 \hour=12 \fi
        \number\hour:\ifnum\minute<10 0\fi\number\minute\the\amorpm}}
\edef\militarytime{\number\hour:\ifnum\minute<10 0\fi\number\minute}
\def\draftlabel#1{{\@bsphack\if@filesw {\let\thepage\relax
   \xdef\@gtempa{\write\@auxout{\string
      \newlabel{#1}{{\@currentlabel}{\thepage}}}}}\@gtempa
   \if@nobreak \ifvmode\nobreak\fi\fi\fi\@esphack}
        \gdef\@eqnlabel{#1}}
\def\@eqnlabel{}
\def\@vacuum{}
\def\draftmarginnote#1{\marginpar{\raggedright\scriptsize\tt#1}}

\def\draft{\oddsidemargin -.5truein
        \def\@oddfoot{\sl preliminary draft \hfil
        \rm\thepage\hfil\sl\today\quad\militarytime}
        \let\@evenfoot\@oddfoot \overfullrule 3pt
        \let\label=\draftlabel
        \let\marginnote=\draftmarginnote
   \def\@eqnnum{(\theequation)\rlap{\kern\marginparsep\tt\@eqnlabel}%
\global\let\@eqnlabel\@vacuum}  }

\def\draft2{
        \def\@oddfoot{\sl preliminary draft \hfil
        \rm\thepage\hfil\sl\today\quad\militarytime}
        \let\@evenfoot\@oddfoot \overfullrule 3pt
        \let\label=\draftlabel
        \let\marginnote=\draftmarginnote
   \def\@eqnnum{(\theequation)\rlap{\kern\marginparsep\tt\@eqnlabel}%
\global\let\@eqnlabel\@vacuum}  }

\def\preprint{\twocolumn\sloppy\flushbottom\parindent 2em
        \leftmargini 2em\leftmarginv .5em\leftmarginvi .5em
        \oddsidemargin -.5in    \evensidemargin -.5in
        \columnsep .4in \footheight 0pt
        \textwidth 10.in        \topmargin  -.4in
        \headheight 12pt \topskip .4in
        \textheight 6.9in \footskip 0pt
        \def\@oddhead{\thepage\hfil\addtocounter{page}{1}\thepage}
        \let\@evenhead\@oddhead \def\@oddfoot{} \def\@evenfoot{} }

\def\numberbysection{\@addtoreset{equation}{section}
        \def\theequation{\thesection.\arabic{equation}}}

\def\underline#1{\relax\ifmmode\@@underline#1\else
        $\@@underline{\hbox{#1}}$\relax\fi}

\def\titlepage{\@restonecolfalse\if@twocolumn\@restonecoltrue\onecolumn
     \else \newpage \fi \thispagestyle{empty}\c@page\z@
        \def\thefootnote{\fnsymbol{footnote}} }

\def\endtitlepage{\if@restonecol\twocolumn \else \newpage \fi
        \def\thefootnote{\arabic{footnote}}
        \setcounter{footnote}{0}}  %\c@footnote\z@ }

\catcode`@=12 \relax

\def\figcap{\section*{Figure Captions\markboth
        {FIGURECAPTIONS}{FIGURECAPTIONS}}\list
        {Figure \arabic{enumi}:\hfill}{\settowidth\labelwidth{Figure
999:}
        \leftmargin\labelwidth
        \advance\leftmargin\labelsep\usecounter{enumi}}}
 \relax
\def\tablecap{\section*{Table Captions\markboth
        {TABLECAPTIONS}{TABLECAPTIONS}}\list
        {Table \arabic{enumi}:\hfill}{\settowidth\labelwidth{Table
999:}
        \leftmargin\labelwidth
        \advance\leftmargin\labelsep\usecounter{enumi}}}
 \relax
\def\reflist{\section*{References\markboth
        {REFLIST}{REFLIST}}\list
        {[\arabic{enumi}]\hfill}{\settowidth\labelwidth{[999]}
        \leftmargin\labelwidth
        \advance\leftmargin\labelsep\usecounter{enumi}}}
 \relax
\makeatletter
\newcounter{pubctr}
\def\publist{\@ifnextchar[{\@publist}{\@@publist}}
\def\@publist[#1]{\list
        {[\arabic{pubctr}]\hfill}{\settowidth\labelwidth{[999]}
        \leftmargin\labelwidth
        \advance\leftmargin\labelsep
        \@nmbrlisttrue\def\@listctr{pubctr}
        \setcounter{pubctr}{#1}\addtocounter{pubctr}{-1}}}
\def\@@publist{\list
        {[\arabic{pubctr}]\hfill}{\settowidth\labelwidth{[999]}
        \leftmargin\labelwidth
        \advance\leftmargin\labelsep
        \@nmbrlisttrue\def\@listctr{pubctr}}}
 \relax
\makeatother

\def\be{\begin{equation}}
\def\ee{\end{equation}}
\def\ba{\begin{eqnarray}}
\def\ea{\end{eqnarray}}

\def\a{\alpha}

\def\g{\gamma}

\def\d{\delta}
\def\D{\Delta}
\def\e{\epsilon}

\def\p{\pi}

\def\Th{\Theta}
\def\m{\mu}
\def\n{\nu}

\def\l{\lambda}

\def\s{\sigma}

\def\no{\noindent}

\def\IR{\relax{\rm I\kern-.18em R}}

\def\bse{\begin{small}\begin{equation*}}
\def\ese{\end{equation*}\end{small}}

\begin{document}
%\draft2

%\renewcommand{\theequation}{\arabic{equation}}
%\renewcommand{\theequation}{\thesection.\arabic{equation}}

\renewcommand{\theequation}{\thesection.\arabic{equation}}
\csname @addtoreset\endcsname{equation}{section}

\newcommand{\eqn}[1]{(\ref{#1})}

\begin{titlepage}
\begin{center}
\strut\hfill
\vskip 1.3cm

%\hfill  [hep-th]\\

\vskip .5in

{\Large \bf Transmission amplitudes from Bethe ansatz equations}

\vskip 0.5in

{\large \bf Anastasia Doikou$^{a}$ and Nikos Karaiskos$^{b}$} \vskip 0.2in

 {\footnotesize $^{a}$Department of Engineering Sciences, University of Patras,\\
GR-26500 Patras, Greece}
\\[2mm]
\noindent
{\footnotesize  $^b$Institut f\"ur Theoretische Physik, Leibniz Universit\"at
Hannover,\\ Appelstra\ss e 2, 30167 Hannover, Germany}

\vskip .1in

%\vskip -.15in

{\footnotesize {\tt E-mail: adoikou@upatras.gr, nikolaos.karaiskos@itp.uni-hannover.de}}\\

\end{center}

\vskip 1.0in

\centerline{\bf Abstract}

We consider the Heisenberg spin chain in the presence of integrable
spin defects. Using the Bethe ansatz methodology, we extract the
associated transmission amplitudes, that describe the interaction
between the particle-like excitations displayed by the models and the
spin impurity. In the attractive regime of the XXZ model, we also
derive the breather's transmission amplitude. We compare our findings
with earlier relevant results in the context of the sine-Gordon model.

\no

\vfill

\end{titlepage}
\vfill \eject

%\def\baselinestretch{1.2}
%\baselineskip 10 pt
%\noindent

\tableofcontents

\section{Introduction}
The presence of impurities in physical systems is an issue of great consequence,
especially when dealing with more realistic models, and confronting experimental data. Along this
spirit, integrability offers a framework where impurities may be naturally
incorporated in a controllable manner. Although there have been numerous recent
advances in both quantum \cite{delmusi}-\cite{annecydef2}, and classical \cite{cozanls}-\cite{doikou-karaiskos-LL}
models with integrable defects, many questions still remain open. In the present investigation, we restrict our
attention on quantum spin chains in the presence of a single integrable defect and extract the physical information concerning
scattering processes within such models, directly from the Bethe ansatz equations.
Our approach thus aims at complementing and enhancing the picture described in earlier works
\cite{corrigan} in the context of integrable field theories.

The algebraic frame describing the presence of a point-like defect in a discrete
integrable theory is by now well established through the quantum inverse scattering
method (QISM) \cite{FT, YBE}. The formulation is based on the existence of a defect Lax operator that satisfies
the same quadratic quantum algebra as the bulk monodromy matrix. In general, let
us consider a one dimensional $(N+1)$-site theory with a point like defect on the
$n$th site. In this case the modified  monodromy matrix of the theory reads as
\be
T(\lambda) = L_{0N+1}(\lambda)\ L_{0N}(\lambda) \ldots \tilde L_{0n}(\lambda-\Theta) \ldots L_{01}(\lambda)\, ,
\label{basic0}
\ee
where $L$ corresponds to the ``bulk'' theory, $\tilde L$ corresponds to the defect
and $\Theta$ is an arbitrary constant corresponding to the ``rapidity'' of the defect.
Both Lax operators satisfy the same quadratic algebra
\be
R_{12}(\lambda_1 -\lambda_2)\ L_1(\lambda_1)\ L_2(\lambda_2) = L_2(\lambda_2)\ L_1(\lambda_1)\ R_{12}(\lambda_1 -\lambda_2)\, ,
\label{basicRLL}
\ee
where the $R$-matrix is a solution of the Yang-Baxter equation (see e.g. \cite{YBE} and references therein).
The monodromy matrix of the theory $T(\l)$, naturally satisfies (\ref{basicRLL}), guaranteeing the
integrability of the model. The Hamiltonian of any generic system with a point-like defect
\be
{\cal H} \propto - \Big ( \sum_{j\neq n=1}^{N+1}  \dot{\check R}_{j j+1}(0) + \dot{\tilde L}_{n+1 n}(0)\
\tilde L^{-1}_{n+1 n}(0) + {\tilde L}_{n+1 n}(0)\ \dot{\check R}_{n-1 n+1}(0)\ \tilde L^{-1}_{n+1 n}(0) \Big ) \label{hamd}
\ee
the ``dot'' denotes the derivative with respect to the spectral parameter. We focus here on the situation where
$L(\lambda) \equiv R(\lambda)$, also define $\check R = {\cal P}\ R$,  ${\cal P}$
is the permutation operator. Recall that the $R$ matrix reduces to the permutation operator at $\lambda =0$.

Note that here we are going to focus on the anti-ferromagnetic regimes of the XXX and XXZ models.
The derivation of the Bethe ansatz equations (BAE) is
straightforward in the case where {\it highest weight} states exist. Amongst others,
the thermodynamic limit of the BAE provides us with the scattering information for
the given model, which we exploit in order to derive our results. Generalization of our
results in the presence of multiple defects is straightforward within the QISM frame.

In the subsequent sections, we investigate the interaction between the particle like
excitations displayed in the XXX and XXZ spin chains and the defect. These
interactions are described by generic transmission matrices, that satisfy the quadratic
algebra (\ref{basicRLL}) with an overall physical factor, which is explicitly computed by means
of the BAE. The XXZ model with a defect is studied in both
the attractive and the repulsive regime. In the latter the formation of bound
states between solitons and anti-solitons, called \textsl{breathers}, is allowed. After
describing the scattering process for the breathers, we derive the transmission
amplitude between a breather and the defect of the theory. We also compare our
findings with earlier results, reaching complete agreement. Finally, the appendices
contain several technical points and physical checks that further confirm our
results.

\section{The isotropic case: XXX model}

We begin our analysis considering the isotropic XXX spin chain in the presence
of a single defect (see also e.g. \cite{AndreiJohannesson}--\cite{zvy} and references therein, 
for both the XXZ and XXZ models). Let us first recall
the quantum Lax operators for the bulk model, $L$, and the for the defect $\tilde L$.
The generic defect matrix is given by
\be
\tilde L(\lambda)  = \begin{pmatrix}
 \lambda + i S^z + {i\over 2}  & i S^- \cr
 iS^+ & \lambda - i S^z + {i\over 2}
\end{pmatrix}, \label{L1}
\ee
where the algebraic objects $S^{z},\ S^{\pm}$ are the generators of the
$\mathfrak{su}_2$ algebra,
\be
\Big [ S^z,\ S^{\pm} \Big ] = \pm S^{\pm}, ~~~~~~ \Big [ S^+,\ S^- \Big ] = 2 S^z.
\ee
The bulk $L$-matrix is actually the $R$-matrix of the model, which corresponds
to the spin-${1\over 2}$ representation of the $\mathfrak{su}_2$, that is the
following identifications are implemented in (\ref{L1}):
\be
S^z \mapsto {\sigma^z \over 2}, ~~~~~~S^{\pm} \mapsto \sigma^{\pm} \, . \label{ident1}
\ee
As usual, $\sigma^z,\ \sigma^{\pm}$ denote the familiar $2\times 2$ Pauli matrices.
In the finite case, for an ${\mathrm n}=2S+1$ dimensional representation of spin $S$, the
algebraic objects $S^z$ and
$\ S^{\pm}$ are represented by ${\mathrm n} \times {\mathrm n}$ matrices defined
as
\be
S^z = \sum_{k=1}^{{\mathrm n}} \alpha_k\ e_{kk}, ~~~~~S^+ = \sum_{k=1}^{{\mathrm n}-1}
C_k\ e_{k k+1}, ~~~~~S^- = \sum_{k=1}^{{\mathrm n}-1} C_k\ e_{k+1 k}\, ,
\label{rep1}
\ee
where we define the matrix elements: $(e_{ij})_{kl} = \delta_{ik}\ \delta_{jl}$ and
\be
\alpha_k = {1\over 2} ({\mathrm n}+1 - 2k), ~~~~~C_k= \sqrt{k ({\mathrm n}-k)}\, .
\ee
This choice may be thought of as the isotropic analogue of the type-II defect studied
in \cite{corrigan}, which will be analyzed in the subsequent sections within the XXZ
spin chain context. The findings of this section may be seen as the quantum discrete
analogues of the results on the Landau-Lifshitz model \cite{doikou-karaiskos-LL}.
Classical scattering in the context of the Landau-Lifshitz model should be considered
and comparison with our findings should provide a more concrete correspondence
between the classical and quantum models in the presence of defects. The Hamiltonian of the model is given by (\ref{hamd})
where $\dot{\tilde L}(0) ={\mathbb I}$,
\be
\dot{\check R}_{j j+1}(0) = {1\over 2} \Big (\sigma_j^x  \sigma_{j+1}^x  + \sigma_j^y  \sigma_{j+1}^y +   \sigma_j^z \sigma_{j+1}^z+ {\mathbb I}_j {\mathbb I}_{j+1} \Big ),
~~~~~ \tilde L_{n+1 n}(0) = {i\over 2}  \Big ( S_n^x  \sigma_{n+1}^x  + S_n^y  \sigma_{n+1}^y + 2S_n^z \sigma_{n+1}^z + {\mathbb I}_n {\mathbb I}_{n+1} \Big ).
\ee

The generic defect matrix as well as the bulk $L$-matrix possess highest weight states ($S^+\ |+\rangle =0$), thus the typical algebraic Bethe variation may be applied and the corresponding Bethe ansatz equations (BAE) are immediately obtained\footnote{The BAE are valid for any $S\neq 0$ real number.}
\be
e_y(\lambda_i-\Theta)\ e_1^N(\lambda_i)= - \prod_{j=1}^N e_2(\lambda_i - \lambda_j), ~~~~~y = 2S,
\ee
where $\Theta$ is the rapidity associated to the defect, and we also define:
\be
e_k(\lambda) = {\lambda +{ik\over 2} \over \lambda - {ik \over 2}}\, .
\ee
Having the Bethe ansatz equations at our disposal, we are now in a position to derive
the physical quantities which describe the scattering processes on the chain.

\subsection{The transmission matrix}
Before we proceed with our analysis we shall recall that in the thermodynamic limit
the solutions of the BAE may be expressed as ``strings'' with a real and an imaginary
part. This is based on the so called ``string hypothesis'', stating that the Bethe roots
may be cast as
\be
\lambda^{(n, j)} = \lambda_0 + {i \over 2}(n+1-2j), ~~~~~j \in \{1, \ldots, n\}\, .
\ee
Recall also that the total spin of a state may be obtained through the asymptotic behavior of the transfer matrix, and is given by the following familiar expression:
\be
S^z = {N \over 2} + S - M\, .
\ee
The spin $S^z$ of the state should be non-negative, thus the restriction $M \leq {N \over 2} +S$
is manifest (below the ``equator''), while the rest of the states can be obtained
by starting from a reference state, which is a ``lowest weight'' state
($S^-\ | - \rangle =0$). The energy and momentum may be also explicitly expressed
in terms of the Bethe roots $\{\lambda_j\}$ (for more details on the Bethe ansatz
formulation and relevant physical expressions see also e.g. \cite{FT, review, doikou-nepo-sun}).

As is well known, the anti-ferromagnetic ground state is a ``filled Fermi sea'' with real solutions, i.e.
1-string configurations. It is easy to check that the ground state has an overall non
zero spin $S^z = \tilde S$, where $\tilde S \equiv S -{1\over 2}$ is the ``shifted''
spin, which naturally emerges through the Bethe ansatz approach, as will be
transparent subsequently when deriving the transmission matrix. Implementation of
suitable string configurations can ``correct'' the spin, i.e. provide a total spin zero,
without modifying the state's energy. In general, it is straightforward to show that in
the presence of an $n$-string the spin becomes
\be
S^z = \tilde S - (n-1) \, ,
\ee
while the energy and the momentum of the configuration are left intact.

A generic state contains $m$ particle-like excitations, which are interpreted as
$m$ holes in the ``filled Fermi sea''. The density of the state may be obtained then,
here we follow the standard formulation \cite{FT, Andrei-Destri} for the derivation.
Starting from the Bethe ansatz equations, and after taking the logarithm and
differentiating, we define the density of the state in the presence of $m$ holes in the
filled Fermi sea as:
\be
\sigma(\lambda) = a_1(\lambda) +{1\over   N} a_y(\lambda- \Theta) - \int_{-\infty}^{\infty}\ d \lambda'\  a_2(\lambda -\lambda')\ \sigma(\lambda') +{1\over N} \sum_{j=1}^m a_2(\lambda -\tilde \lambda_j) \, ,
\ee
where we have defined
\be
a_n(\lambda) = {i \over 2 \pi} {d \over d\lambda} \ln \Big (e_n(\lambda)\Big) \, .
\label{defan34}
\ee
Passing to the thermodynamic limit, we have exploited the following basic
formula in the presence of $m$ holes, with associated rapidities $\tilde \lambda_j$:
\be
{1\over N}\sum_{j=1}^M f(\lambda_j)  \to \int_{-\infty}^{\infty}\ d\lambda\ f(\lambda)\ \sigma(\lambda)- {1\over N} \sum_{j=1}^{m} f(\tilde \lambda_j)\, . \label{basic1}
\ee
The Fourier transform of $a_n$ is needed subsequently, and is
given by\footnote{We have considered the following conventions for the Fourier transformations:
\be
f(\lambda) = {1\over 2 \pi} \int d\omega\ e^{-i\omega \lambda} \hat f(\omega)\, , ~~~~~\hat f(\omega) = \int d\lambda\ e^{i\omega \lambda} f(\lambda)\, .
\ee}
\be
\hat a_n(\omega) = e^{-n{|\omega| \over 2}}.
\ee

For our purposes here, we shall focus on the case where $m=2$ ($N$ is assumed to
be odd), in order to derive the ``kink'' scattering matrix, as well as the transmission
matrix. Using the machinery described above, the density may be expressed in a
compact form as
\be
\sigma(\lambda) = \sigma_0(\lambda) + {1\over N} \Big ( \sum_{k=1}^2r_s(\lambda -\tilde \lambda_k) + r_t(\lambda-\Theta) \Big )\, , \label{dens1}
\ee
with $\s_0(\l)$ being the density of the ground state. The Fourier transforms of the
latter quantities have been computed
\be
\hat \sigma_0(\omega)  = {1\over 2 \cosh ({\omega \over 2})}\, , ~~~~~
\hat r_s(\omega)= {e^{-{|\omega| \over 2}} \over 2 \cosh ({\omega \over 2})}\, ,
~~~~~~\hat r_t(\omega) = {e^{-(y-1){|\omega| \over 2}} \over 2 \cosh ({\omega \over 2})}\, .
\ee
Recall also that
\be
\sigma_0(\lambda) = \varepsilon(\lambda), ~~~~\mbox{and}~~~~~~ \varepsilon(\lambda) = {1\over 2 \pi}{d p(\lambda)\over d \lambda} \, ,\label{EP}
\ee
with $\varepsilon$ and $p$ being the energy and the momentum of the hole
excitation (kink), respectively. Recall also that
\be
\sigma(\lambda) = {1\over N}{ d h(\lambda) \over d\lambda}
\ee
$h(\lambda)$ is the so-called counting function and $h(\tilde \lambda_i) = J_i$, where $J_i$ are integer numbers.

In order to identify the scattering amplitude between two holes as well as the hole-defect transmission amplitude, we compare the expression providing the density of
the state (\ref{dens1}) with the so called quantization condition, with respect to the
hole with rapidity $\tilde \lambda_1$:
\be
\Big (e^{i N p(\tilde \lambda_1)}\ S(\tilde \lambda_1, \tilde \lambda_2, \Theta) -1 \Big )|\tilde \lambda_1,\ \tilde \lambda_2\rangle =0 \, ,
\label{QC1}
\ee
where $S=e^{i\Phi}$, and $p(\tilde \lambda_1)$ is the momentum of the respective hole.
Comparison of the quantization condition with the state's density (\ref{dens1}) would
provide the ``kink-kink'' scattering amplitude as well as the transmission amplitude,
given that the factorization of the scattering is evident (see also \cite{doikou-nepo-mezi2}). More precisely, the study of the one-hole state would simply provide the
transmission amplitude, which physically describes the interaction between the
particle-like excitation and the defect, thus factorization of the type: $S(\tilde \lambda_1, \tilde \lambda_2, \Theta) = S_s(\tilde \lambda_1, \tilde \lambda_2)\ T(\tilde \lambda_1, \Theta)$,
in the case of the two-hole state is manifest. Keeping these considerations in mind,
the kink-kink amplitude as well as the transmission amplitude for the XXX model with
a single defect can be derived then as
\be
S_s(\lambda) = \exp \Big [ - \int_{-\infty}^{\infty}\ {d\omega \over \omega} e^{-i \omega \lambda} \hat r_s(\omega)\Big ], ~~~~~~
T(\hat \lambda) = \exp \Big [ - \int_{-\infty}^{\infty}\ {d\omega \over \omega} e^{-i \omega \hat \lambda} \hat r_t(\omega) \Big ] \, ,
\label{ST}
\ee
where $\lambda \equiv \tilde \lambda_1 - \tilde \lambda_2$ and
$\hat \lambda \equiv \tilde \lambda_1 -\Theta$.

Taking now into account the following useful identity
\be
{1\over 2}\int_0^{\infty}\ {d\omega \over \omega} {e^{-{\mu \omega \over 2}} \over \cosh{\omega \over 2}} =
\ln {\Gamma({\mu +1 \over 4}) \over \Gamma({\mu +3 \over 4}) }\, ,
\label{use1}
\ee
the familiar expression for the XXX kink-kink scattering amplitude \cite{FT, Kulish-Resh} is reproduced
in terms of $\Gamma$-functions
\be
S_s(\lambda) =  {\Gamma(-{i\lambda \over 2} + {1\over 2})\ \Gamma( {i\lambda \over 2} +1) \over \Gamma(-{i\lambda \over 2} +1)\ \Gamma({i\lambda \over 2} + {1\over 2})}\, .
\ee
This is the first eigenvalue of the scattering matrix and it is three-fold degenerate
due to the underlying $\mathfrak{su}_2$ symmetry. There is one more eigenvalue
corresponding to the singlet state, which may be derived by considering the state
with two holes and one 2-string with a real center:
\be
\lambda_0 = {\tilde \lambda_1 + \tilde \lambda_2 \over 2}\, .
\ee
The second eigenvalue may be identified then, and turns out to be
\be
{S_2(\lambda) \over S_s(\lambda)} = {i\lambda -1 \over i\lambda +1}\, .
\ee
The $S$-matrix, which satisfies the Yang-Baxter equation, may be then
cast as:
\ba
{\mathbb S}(\lambda)  = {S_s(\lambda)\over i\lambda+1} \begin{pmatrix}
 i\lambda+1 & & & \cr
            &i\lambda  &1 & \cr
            &1 &i\lambda \ &\cr
            &&&  i\lambda+1
\end{pmatrix}, \label{Smatrix1}
\ea
which is the familiar XXX scattering matrix \cite{FT}. Unitarity and
crossing are also explicitly checked and are satisfied by the extracted $S$-matrix (see
Appendix A).

The transmission factor may be also identified through expression (\ref{ST}), and is
found to be given by
\be
T(\hat \lambda, \tilde S) =  {\Gamma({i\hat\lambda \over 2}+{\tilde S\over 2} + {3\over 4})\ \Gamma(- {i\hat\lambda \over 2} +{\tilde S\over 2}+ {1\over 4}) \over \Gamma({i\hat\lambda \over 2} +{\tilde S\over 2}+{1\over 4})\ \Gamma(-{i\hat\lambda \over 2} +{\tilde S \over 2}+ {3\over 4})}\, .
\label{exp1}
\ee
The spin associated to this state is identified through the BAE, and is given as: $S^z = \tilde S + {1\over 2},\ (\tilde S = S -{1\over 2})$, corresponding to the highest spin $S^z$ eigenvalue (Appendix
A). Notice the ``shift'' of the spin which naturally emerges through the Bethe ansatz
process. We have been able so far to identify the first eigenvalue of the transmission
matrix. As shown in the Appendix A, there are only two distinct $(\tilde {\mathrm n}=2 \tilde S +1)$-fold degenerate eigenvalues associated to the generic
transmission matrix.

As already pointed out, in order to identify the transmission amplitude, it is sufficient
to consider the state with one hole ($N$ even) (see also e.g. \cite{doikou-nepo-mezi1}). The other distinct eigenvalue may be found using suitable string configurations that modify the spin, but leave the energy of the sate intact. Considering $n$-strings,
suitably positioned with respect to $\tilde \lambda_1$ and $\Theta$, one may determine each eigenvalue corresponding to the appropriate spin $S^z$ eigenvalue $S^z \in \Big \{S_{min}^z, \ldots , \tilde S+1/2 \Big \}$. The rest of the negative
spins may be obtained starting from a reference state, which is a ``lowest'' weight state (see also Appendix A). For instance, it is easy to identify the spin $S^z$ eigenvalue of a state with one hole of rapidity $\tilde \lambda_1$, and one $n$-string: $n \in \Big \{2, \ldots, n_0\Big \}$, where we define
\ba
&& n_0 = \tilde S +{3\over 2}\ \Rightarrow \ S_{min}^z = 0, ~~~\mbox{$\tilde S$ half-integer}\cr
&& n_0 = \tilde S +1\ \Rightarrow \ S_{min}^z = {1\over 2}, ~~~\mbox{$\tilde S$ integer}.\
\ea
The spin turns out then to be
\be
S^z = \tilde S + {1\over 2} -(n-1)\, .
\label{spin-string}
\ee
The real center of the $n$-string is positioned at
\be
\lambda_0 = {x-1 \over x }\tilde \lambda_1 + {\Theta \over x}, ~~~~~x = {2S \over n-1}\, .
\ee
The position of the $n$-string is determined using suitable quantum numbers, which
characterize the state. We shall not give the technical details of the proof here,
however the interested reader is referred to \cite{doikou-nepo-sun} for more details.
Given the arguments above, the second eigenvalue of the transmission matrix can be
derived for such a state:
\be
{T_2(\hat \lambda, \tilde S) \over T(\hat \lambda, \tilde S)} = {i \hat \lambda -\tilde S - {1\over 2} \over i \hat \lambda+\tilde S + {1\over 2}}\, ,
\label{eigen_ratio}
\ee
corresponding to the spin eigenvalue (\ref{spin-string}).

Having the eigenvalues of the transmission matrix at hand, and bearing in mind that
for purely transmitting defects the following quadratic algebra is satisfied
\cite{delmusi}
\be
{\mathbb S}_{12}(\lambda_1 -\lambda_2)\ {\mathbb T}_1(\lambda_1)\ {\mathbb T}_2(\lambda_2) = {\mathbb T}_2(\lambda_2)\ {\mathbb T}_1(\lambda_1)\ {\mathbb S}_{12}(\lambda_1 -\lambda_2)\, ,
\label{rttb}
\ee
 where ${\mathbb S}$ is given in (\ref{Smatrix1}), we conclude that
\ba
{\mathbb T}(\hat \lambda, \tilde S)  = {T(\hat \lambda, \tilde S) \over i \hat \lambda + \tilde S + {1 \over 2}} \begin{pmatrix}
 i\hat \lambda + S^z + {1\over 2}  & S^- \cr
       S^+     & i \hat \lambda - S^z + {1\over 2}
\end{pmatrix}. \label{exp2}
\ea
This is the generic matrix associated this time to the spin $\tilde S = S - {1\over 2}$
representation of $\mathfrak{su}_2$. The ``shift'' of the spin
via the Bethe ansatz process, is once more pointed out.

The findings of this paragraph can be further confirmed by checking the basic
requirements of unitarity and crossing symmetry, and of course by recalling  that the
transmission matrix has to satisfy (\ref{rttb}). The transmission matrix presented
above clearly satisfies the fundamental algebraic relation due to its structure.
Unitarity and crossing symmetry have been also checked by inspection and confirmed
for the transmission matrix (\ref{exp2}) (see Appendix B for details). In addition to
the derivation of the transmission matrix eigenvalues via the BAE, there exist strong
physical and algebraic arguments, as well as various checks regarding the validity of
(\ref{exp2}) as is manifest from the discussion above. It is also worth noting that this
is the first time to our knowledge that such expressions are computed in the context
of the XXX spin chain with a defect from the Bethe ansatz equations, and expressions
(\ref{exp1}), (\ref{exp2}) are as far as we can tell novel.

\section{The anisotropic case: XXZ model}
Our investigation on the defects is carried on with the anisotropic XXZ spin chain.
Based on the analysis of this model we shall be able to make contact with relevant
results extracted in the context of the sine-Gordon model. We shall focus in the
present study on the so called type-II defects \cite{corrigan}. Let us first introduce the
general defect $\tilde L$-matrix \cite{kulres}:
\be
\tilde L(\lambda)  = \begin{pmatrix}
 \sinh\Big (\mu (\lambda + iS^z +{i\over 2})\Big )  & \sinh (i\mu)\ S^- \cr
\sinh(i\mu)\ S^+ & \sinh \Big ( \mu (\lambda -iS^z + { i\over 2}) \Big)
\end{pmatrix} \, ,
\label{ldefect}
\ee
where we define $q = e^{i\mu}$, measuring the anisotropy of the model as
$\D \equiv \cosh i\m$. The bulk $L$-matrix is basically the $R$-matrix of the XXZ
model, that is the spin $1 \over 2$ representation of the latter expression, i.e. one
simply implements in (\ref{ldefect}) the identifications (\ref{ident1}).

The $\tilde L$-matrix is the typical solution of the quadratic algebra associated to $\mathfrak{U}_q(\mathfrak{sl}_2)$, which reads as:
\be
\Big [S^{z},\ S^{\pm}\Big ] = \pm S^{\pm}\, , ~~~~\Big [S^+,\ S^- \Big ] = {q^{2S^z} - q^{-2S^z} \over q -q^{-1}}\, . \label{qalg}
\ee
In the finite case, which will be considered here, the generators are represented by ${\mathrm n} \times {\mathrm n}$  (${\mathrm n}=2S+1$), matrices as:
\be
S^z = \sum_{k=1}^{{\mathrm n}} \alpha_k\ e_{kk}\, , ~~~~~S^+ = \sum_{k=1}^{{\mathrm n}-1} \tilde C_k\ e_{k k+1}\, , ~~~~~S^- = \sum_{k=1}^{{\mathrm n}-1} \tilde C_k\ e_{k+1 k}\, ,
\ee
where we define
\be
\alpha_k = {1\over 2} ({\mathrm n}+1 - 2k)\, , ~~~~~\tilde C_k= \sqrt{[k]_q [{\mathrm n}-k]_q}\, , ~~~~~~[x]_q = {q^x - q^{-x} \over q -q^{-1}}\, .
\ee
Note that in \cite{corrigan} the $\tilde L$-matrix (\ref{ldefect}) is also used, but an infinite dimensional representation is employed; nevertheless, the structure of the defect matrices as well as the extracted physical quantities presented in \cite{corrigan} are similar to our expressions, as will be evident later in the text. The Hamiltonian of the model is given by (\ref{hamd})
where $\dot{\tilde L}(0) = \mu\ diag\Big ( \cosh i\mu(S^z+{1\over 2}), \ \cosh i\mu(-S^z+{1\over 2})\Big )$,
\ba
&&\dot{\check R}_{j j+1}(0) = {\mu \over 2} \Big (\sigma_j^x  \sigma_{j+1}^x  + \sigma_j^y  \sigma_{j+1}^y +  \cosh (i\mu) \sigma_j^z \sigma_{j+1}^z + \cosh(i\mu){\mathbb I}_j {\mathbb I}_{j+1} \Big )\nonumber\\
&&\tilde L_{n+1 n}(0) = {\sinh(i\mu )\over 2} (S_n^x  \sigma_{n+1}^x  + S_n^y  \sigma_{n+1}^y )   +\cosh({i\mu \over 2})\sinh (i \mu S_n^z) \sigma_{n+1}^z +\sinh({i\mu \over 2}) \cosh(\i\mu S_n^z) {\mathbb I}_{n+1}. \nonumber
\ea

The generic defect matrix and the bulk $L$-matrix possess highest weight states, thus as in the isotropic case the typical algebraic Bethe variation may be applied and the corresponding Bethe ansatz equations are immediately obtained\footnote{The BAE are valid for any $S \neq 0$ real number.} having the following standard form
\be
e_y(\lambda_i-\Theta)\ e_1^N(\lambda_i)= - \prod_{j=1}^N e_2(\lambda_i - \lambda_j)\, , ~~~~~y=2S\, ,
\ee
with $\Th$ being again the rapidity of the defect, and we also introduce the notation
\be
e_n(\lambda) = {\sinh (\mu(\lambda +{in\over 2})) \over \sinh(\mu (\lambda - {in \over 2}))}\, .
\ee
We may now proceed with the derivation of the transmission matrix distinguishing
two regimes, the repulsive and the attractive, depending on the value of the
coupling constant. Bear in mind that we wish to compare our findings with similar
results in the context of the sine-Gordon model. Therefore, it is useful to provide the
relation between the sine-Gordon coupling constant $\beta^2$, and the anisotropy parameter $\mu$ of the XXZ model (see also e.g. \cite{doikou-nepo-breather} for a more detailed discussion):
\ba
\beta^2 &=& 8(\pi - \mu)\, ,~~~~~4\pi < \beta^2 < 8 \p ~~~~~~~~~ \mbox{repulsive regime}, \cr
\beta^2 &=& 8\mu\, ,  ~~~~~~~~~~~~~~0< \beta^2 < 4\pi ~~~~~~~~~~\mbox{attractive regime}.
\ea
Note that in the attractive regime the formulation of bounds states between solitons
and anti-solitons of zero spin (topological charge), the so called ``breathers'' is
allowed.

\subsection{The repulsive regime; the transmission matrix}
We shall first consider the repulsive regime, and derive the corresponding transmission
matrix. In this regime the antiferromagnetic ground state is a filled Fermi sea with real strings ($N$
odd), as in the isotropic case. For our purposes, it suffices to consider here the state
with two holes. The density associated to this state is obtained in the thermodynamic
limit by following the logic described in the previous section
\be
\sigma(\lambda) = a_1(\lambda) + {1\over N} a_y(\lambda-\Theta) - \int_{-\infty}^{\infty}\ d\lambda'\ a_2(\lambda -\lambda')\ \sigma(\lambda') + {1\over N}\sum_{j=1}^2 a_2(\lambda - \tilde \lambda_j)\, ,
\ee
where the basic formula in  the presence of two ``holes'' of rapidities $\tilde \lambda_j$  (\ref{basic1}) was exploited. Note that $a_n(\l)$ is defined again
as in (\ref{defan34}), whereas its Fourier transformation is given by
\be
\hat a_n(\omega) = {\sinh ((\nu - n) {\omega \over 2}) \over \sinh  ({\nu \omega \over 2})}\, , ~~~~\nu = {\pi \over \mu}, ~~~~0 < n < 2\nu\, .
\label{afour}
\ee
In the present subsection, we restrict ourselves to the values $0 < y < 2\nu$. Results
on generic values of $y$ are presented in Appendix C.

As in the isotropic case, the density may be expressed in a compact form as
\be
\sigma(\lambda) = \sigma_0(\lambda) + {1\over N} \Big (\sum_{k=1}^2r_s(\lambda -\tilde \lambda_k) + r_t(\lambda-\Theta) \Big )\, . \label{dens2}
\ee
The Fourier transforms of the latter quantities have been explicitly determined
\be
\hat \sigma_0(\omega)  = {1\over 2 \cosh ({\omega \over 2})}\, , ~~~~~
\hat r_s(\omega)= {\sinh ((\nu-2){\omega \over 2})\over  2\sinh((\nu -1){\omega \over 2}) \cosh ({\omega \over 2})}\, ,
~~~~~~\hat r_t(\omega) = {\sinh ((\nu-y ){\omega \over 2})\over 2\sinh((\nu -1){\omega \over 2}) \cosh ({\omega \over 2})}\, .
\ee
Relations (\ref{EP}) for the energy and momentum of the particle-like (``soliton'') excitation are also valid here.

In order to identify the scattering amplitude between two holes, we compare the
expression providing the density of the state with the quantization condition
(\ref{QC1}), with respect to the excitation of rapidity $\tilde \lambda_1$. We are now
in the position to compute the soliton-soliton scattering amplitude, as well as the
transmission amplitude for the XXZ model using the expressions (\ref{ST}). Taking
into account the following useful identity
\be
{1\over 4} \int_0^{\infty}\ {dx \over x}\ {e^{-\mu x} \over \sinh x\ \sinh \beta x} = \ln \prod_{k=0}^{\infty} \Gamma({\mu \over 2} + {\beta \over 2} + k \beta +{1\over 2}) \, ,
\label{use2}
\ee
we may easily reproduce the well known expression for the sine-Gordon soliton-soliton scattering (see also \cite{zamo, corrigan}),
\ba
S_s(\lambda, \gamma) & = & \prod_{k=0}^{\infty}
{\Gamma(z +2(k+1)\gamma)\ \Gamma(z + 2k \gamma +1) \over \Gamma(z +(2k+1)\gamma)\ \Gamma(z + (2k +1)\gamma +1)}  \cr
& \times & \,
{\Gamma(-z +(2k+1)\gamma)\ \Gamma(-z + (2k+1) \gamma +1) \over \Gamma(-z +2(k+1)\gamma)\ \Gamma(-z + 2k \gamma +1) } \, ,
\label{S2}
\ea
where we now define
\be
z =  i \gamma \lambda\, , ~~~~~\gamma = {1\over \nu-1}\, .
\ee
Suitable configurations corresponding to soliton/anti-soliton states may be
formulated, so that all the $S$-matrix eigenvalues may be identified. We shall not
give the details of such a derivation here, however we refer the interested reader to
e.g. \cite{doikou-nepo-breather, doikou-nepo-critical} and references therein for a
more detailed analysis. The $S$-matrix, solution of the Yang-Baxter equation, is
structurally similar to the ``bare'' $R$-matrix, and is given by
\ba
{\mathbb S}(\lambda)  = {S_s(\lambda, \gamma) \over a(\lambda, \gamma)}\begin{pmatrix}
 a(\lambda,\gamma)  & & & \cr
            &b(\lambda,\gamma)&c(\gamma)& \cr
            &c(\gamma)&b(\lambda, \gamma)&\cr
            &&&a(\lambda,\gamma)
\end{pmatrix} \, ,
\label{Smatrix2}
\ea
where
\be
a(\lambda, \gamma) = \sin( \pi \gamma (i\lambda + 1))\, ,  ~~~~\beta(\lambda,\gamma) =\sin( i\pi \gamma\lambda)\, , ~~~~c(\gamma) = \sin( \pi \gamma)\, .
\ee
Note that the $S$-matrix is essentially a renormalized $R$-matrix, given that both
the spectral parameter as well as the anisotropy parameter are renormalized, as is
clear from the expressions above.  The $S$-matrix is in agreement with the results
on the sine-Gordon model (see also \cite{zamo, corrigan, doikou-nepo-breather, doikou-nepo-critical}).

Recalling expressions (\ref{ST}) we can also extract the transmission amplitude. Based
on the state described above, as well as the factorization argument, we find the first
eigenvalue of the transmission matrix
\ba
T(\hat \lambda, \gamma, \tilde S) &=&  \prod_{k=0}^{\infty}
{\Gamma(\hat z +\gamma \tilde S +{\gamma \over 2}+(2k+1)\gamma)\ \Gamma(\hat z -\gamma \tilde S- {\gamma \over 2}+(2k+1)\gamma+1)\over \Gamma(\hat z +\gamma \tilde S +{\gamma \over 2}+2k\gamma)\ \Gamma(\hat z -\gamma \tilde S-{\gamma \over 2} +2(k+1)\gamma+1)} \cr
&\times & {\Gamma(-\hat z +\gamma \tilde S +{\gamma \over 2}+2k\gamma)\ \Gamma(-\hat z - \gamma \tilde S-{\gamma \over 2} +2(k+1)\gamma+1)\over \Gamma(-\hat z + \gamma \tilde S+{\gamma \over 2}+(2k+1)\gamma)\ \Gamma(-\hat z - \gamma \tilde S-{\gamma \over 2} +(2k+1)\gamma+ 1)}\, , \label{expb1}
\ea
where $\hat z = i \hat\lambda \gamma$ and $\tilde S = S- {1\over 2}$ is the ``shifted''
spin. As already mentioned, in order to derive the transmission amplitude it is
sufficient to consider the state with one hole. In this case, the corresponding spin
eigenvalue is computed explicitly and turns out to be
\be
S^z = {\nu \over \nu -1}(\tilde S + {1\over 2})\, .
\ee
One immediately observes an overall renormalization factor ${\nu \over \nu -1}$,
which is equal to the ratio of the bare anisotropy parameter $\nu$ over the
renormalized one $\nu-1$; the physical (renormalized) spin reduces then to the
expected one: $S_{ph.}^z = \tilde S + {1\over 2}$.

Alongside the $S$-matrix (\ref{Smatrix2}), the transmission matrix ${\mathbb T}$
satisfies the quadratic algebra (\ref{rttb}). We conclude that the transmission matrix
may be cast then as
\ba
{\mathbb T}(\hat \lambda, \gamma, \tilde S)  ={ T(\hat \lambda, \gamma, \tilde S) \over \sin( \pi\gamma (i\hat \lambda + \tilde S + {1\over 2}))} \begin{pmatrix}
 \sin (\pi \gamma(i \hat \lambda + S^z + {1\over 2}))  & \sin( \pi \gamma)\ S^-\cr
  \sin( \pi \gamma)\ S^+  &  \sin( \pi \gamma(i \hat \lambda -S^z + {1\over 2}))
\end{pmatrix}\, ,
\label{TT}
\ea
where $q^{S^z},\ S^{\pm}$ correspond to the spin $\tilde S$ representation of
$\mathfrak{U}_q(\mathfrak{sl}_2)$, where now $q = e^{i\pi \gamma}$. We should point out that
the transmission matrix is essentially a renormalized defect matrix, as is manifest
from the structure of ${\mathbb T}$ and $\tilde L$-matrices. Having said this,
it is straightforward to verify that ${\mathbb T}$ is indeed a solution of the
fundamental algebra (\ref{rttb}). Moreover, based on the analysis presented in
Appendix A, we have checked by inspection that the transmission matrix satisfies
unitarity and crossing symmetry, hence its validity is completely confirmed. As in the
isotropic case, appropriate string configurations suitably positioned with respect to
$\tilde \lambda_1,\ \Theta$ provide the various eigenvalues of the transmission
matrix. However, in the trigonometric case this is a highly intricate task, and will be
left for separate investigations. It is also worth noting that this is the first time as far as we know that such expressions are
computed via the Bethe ansatz formulation in the trigonometric case. A detailed comparison with
earlier results on the transmission matrix of the sine-Gordon model will be given in the next subsection.

\subsection{The attractive regime}
\subsubsection{The soliton transmission matrix}

We shall now focus on the attractive regime; in this regime bound states
between solitons and anti-solitons, the so-called ``breathers'', exist. Thus the
scattering between the breathers and the defect may be also investigated. As was
shown in earlier studies, the ground state in the attractive regime consists of the so-called
negative parity strings (see also e.g. \cite{doikou-nepo-breather} and references therein)
\be
\lambda^{(-)} = \lambda + {i\pi \over 2 \mu}\, .
\ee
The BAE are modified then as follows
\be
g_y(\lambda_i-\Theta)\ g_1^N(\lambda) = - \prod_{j=1}^M e_2(\lambda_i - \lambda_j)\, ,
\ee
where we define
\be
g_n(\lambda)={\cosh (\mu(\lambda +{in\over 2})) \over \cosh(\mu (\lambda - {in \over 2}))}\, .
\ee

A generic state with two particle excitations (two holes in the filled Fermi sea of
negative parity strings) is considered and the density associated to this state may be
derived based on the standard formulation \cite{FT, Andrei-Destri}. It turns out that the derived state density is given by the following expression
\be
-\sigma(\lambda) = b_1(\lambda) + {1\over N} b_y(\lambda-\Theta) - \int_{-\infty}^{\infty}\ d\lambda'\ a_2(\lambda -\lambda')\ \sigma(\lambda') + {1\over N}\sum_{j=1}^2 a_2(\lambda - \tilde \lambda_j) \, ,
\ee
where the formula (\ref{basic1}) in the presence of two ``holes'' of rapidities $\tilde \lambda_j$  has been exploited. The Fourier transformation of $a_n$ is given in
(\ref{afour}), whereas the Fourier transform for $b_n$ is found to be
\ba
\hat b_n(\omega) &=& - {\sinh (  {n \omega \over 2}) \over \sinh  ({\nu \omega \over 2})}\, , ~~~0 < n < \nu\, , \cr
\hat b_n(\omega) &=& - {\sinh ( (n- 2\nu) {\omega \over 2}) \over \sinh  ({\nu \omega \over 2})}\, , ~~~\nu <n < 2\nu\, .
\label{bfour}
\ea
For the sake of simplicity, we consider here the case $0 < y < \nu$. Results
regarding generic values of $y$ are presented in Appendix C. Similarly to the previous
sections, the density is compactly expressed as
\be
\sigma(\lambda) = \sigma_0(\lambda) + {1\over N} \Big (\sum_{k=1}^2 r_s(\lambda-\tilde \lambda_k) + r_t(\lambda-\Theta) \Big )\, , \label{dens3}
\ee
whereas the Fourier transforms of the latter quantities are given by
\ba
&&\hat \sigma_0(\omega)  = {1\over 2 \cosh ((\nu -1){\omega \over 2})}\,, ~~~~~
 \hat r_s(\omega)= -{\sinh ((\nu-2){\omega \over 2})\over  2\sinh({\omega \over 2}) \cosh ((\nu -1){\omega \over 2})}\, , \cr
&& \qquad\qquad\qquad \hat r_t(\omega) = {\sinh (y {\omega \over 2})\over 2\sinh({\omega \over 2}) \cosh ((\nu-1){\omega \over 2})}\, .
\ea
Note also that relations (\ref{EP}) also hold for the energy and momentum of the
particle-like excitation.

Comparison of the expression providing the density of the state (\ref{dens3}) with the quantization condition (\ref{QC1}), with respect to the excitation of rapidity $\tilde \lambda_1$, leads to the derivation of the soliton-soliton scattering as well as the
transmission amplitude, given by (\ref{ST}). Taking into account the identity
(\ref{use2}) we reproduce the well known expression for the sine-Gordon soliton-soliton scattering in the attractive regime (compare for instance with the notation
used in \cite{corrigan}), which is given by (\ref{S2}), but now we define
\be
z = i \lambda\, , ~~~~~\gamma = \nu-1\, .
\ee
Soliton anti-soliton configurations leading to all the $S$-matrix eigenvalues may be identified \cite{doikou-nepo-breather, doikou-nepo-critical}. The $S$-matrix, solution of the Yang-Baxter equation as well, is structurally similar to the ``bare'' $R$-matrix, and is given as (\ref{Smatrix2}),
where we now define
\be
a(\lambda, \gamma) = \sin \pi(i\lambda +  \gamma)\, ,  ~~~~\beta(\lambda,\gamma) =\sin i\pi\lambda\, ,
~~~~c(\gamma) = \sin \pi \gamma\, .
\ee

In a similar fashion, we also identify the transmission matrix, which describes the
interaction between the soliton and the defect. Based on the state described above, as
well as the factorization argument, we extract the first eigenvalue of the transfer
matrix, which is expressed as
\ba
T(\hat \lambda, \gamma) &=& \prod_{k=0}^{\infty}
{\Gamma(\hat z-\xi+2(k+1)\gamma +{1\over 2})\ \Gamma(\hat z+\xi+2k\gamma +{1\over 2})
\over \Gamma(\hat z-\xi+(2k+1)\gamma +{1\over 2})\ \Gamma(\hat z+\xi+(2k+1)\gamma +{1\over 2})}\cr &\times&
{\Gamma(-\hat z-\xi+(2k+1)\gamma +{1\over 2})\ \Gamma(-\hat z+\xi+(2k+1)\gamma +{1\over 2})
\over \Gamma(-\hat z-\xi+2(k+1)\gamma +{1\over 2})\ \Gamma(-\hat z+\xi+2k\gamma +{1\over 2})}\, , \label{factor3}
\ea
where
\be
\hat z = i \hat \lambda\, , ~~~~~~\xi = S + {\gamma \over 2}\, .
\ee
The transmission matrix ${\mathbb T}$ satisfies the quadratic algebra (\ref{rttb})
together with the $S$-matrix derived above. We conclude that the transmission
matrix ${\mathbb T}$ may be then cast as (set ${i\hat \lambda \over \gamma} = iu$)
\ba
{\mathbb T}(u, \gamma, \tilde S)  = { T(u, \gamma, \tilde S) \over \sin( \pi \gamma (i u +\tilde S+ {1\over 2}))} \begin{pmatrix}
\sin (\pi \gamma(i u + S^z + {1\over 2}))  & \sin( \pi \gamma)\ S^-\cr
\sin( \pi \gamma)\ S^+  &  \sin( \pi \gamma(i u -S^z + {1\over 2}))
\end{pmatrix}\, .
\label{TT2}
\ea
The elements $S^z,\ S^{\pm}$ form now the spin $\tilde S =0$ representation of
$\mathfrak{U}_q(\mathfrak{sl}_2)$ with $q^{i\pi \gamma}$, which is an infinite dimensional
representation. Notice that here we have used the fact that:\\
$\sin (\pi\gamma (i u + {1\over 2}-{S+{1\over 2} \over \gamma})) = \pm \sin( \pi \gamma(iu + {1\over 2}))$ or
$\pm \cos( \pi \gamma (iu + {1\over 2}))$ given that
$S + {1\over 2}$ is an integer or half-integer.
Computation of the spin of the state with one hole via the Bethe
ansatz equations leads to $S^z = {\n \over 2}$, i.e. the renormalized spin is
${1\over 2}$ (${\nu}$ is the renormalization factor in the attractive regime),
which is basically the spin of the hole, denoting that the defect spin is effectively
zero, confirming the relevant result ($\tilde S =0$) through the derivation of the transmission matrix.

To efficiently compare with earlier relevant results from the Sine-Gordon model
\cite{corrigan}, it is convenient to introduce some useful notation; first we shift
$\lambda$ such that
\be
i\hat \lambda \to i \hat \lambda + \Lambda,
\ee
where $\Lambda$ is an arbitrary constant. Also define the following quantities:
\ba
z_j = -i \hat\lambda - {i\gamma \over \pi} \eta_j,
~~~~\eta_1 = {i\pi \over \gamma} ( \Lambda +\xi), ~~~~~\eta_2 = {i \pi \over \gamma} (\Lambda-\xi ), ~~~~j \in \{1,\ 2\}. \label{ident2} \cr
\ea

The transmission factor is then expressed in terms of $z_1,\ z_2$ as
\be
T(z_1, z_2) = {\sin \pi(z_2 + {1\over 2}) \over \pi}\ \rho_d(z_1, z_2)\, ,
\ee
where the overall physical factor $\rho_d$ is given by
\ba
\rho_d(z_1, z_2) &=& \Gamma({1\over 2} -z_1)\ \Gamma({1\over 2}-z_2) \prod_{k=1}^{\infty}
{\Gamma(z_1 +(2k-1)\gamma+{1\over 2})\ \Gamma(z_2 +(2k-1)\gamma+{1\over 2})\over \Gamma(z_1 +2k\gamma+{1\over 2})\ \Gamma(z_2 +2k\gamma+{1\over 2})} \cr
&& \qquad \times  ~{\Gamma(-z_1 +2k\gamma+{1\over 2})\ \Gamma(-z_2 +2k\gamma+{1\over 2})\over \Gamma(-z_1 +(2k-1)\gamma+{1\over 2})\ \Gamma(-z_2 +(2k-1)\gamma+{1\over 2})}\, .
\ea
The latter expression is identical to the one extracted in \cite{corrigan} for the type-II
defects, taken into account the identifications (\ref{ident2}). Having determined the
soliton transmission matrix we now proceed with the derivation of the breather
transmission matrix.

\subsubsection{The breather transmission amplitude}

The breathers are in general identified within the Bethe ansatz frame by suitable string configurations. To fully describe this scattering process for the breathers it is necessary to take into consideration two sets of Bethe ansatz equations; the first set describes the negative parity one-strings, while the second one describes the breather itself. The second set of BAE is necessary in order to derive the energy and momentum of the breather, and also compare with the quantization condition with respect to the breather (for more details on breathers and their interactions we refer the interested breather to \cite{doikou-nepo-breather} and references therein).

A state with two light breathers with rapidities $\bar \lambda_1,\ \bar \lambda_2$ will be considered. We shall basically deal with the lightest breathers for simplicity; a generalization of the results concerning higher breathers is then straightforward \cite{doikou-nepo-breather}, and is given at the end of the subsection. The lightest breather is described by one positive parity (real) string with rapidity $\bar \lambda_i$, then the BAE for the state with two breathers are expressed as :
\be
g_y(\lambda_i -\Theta)\ g_1^N(\lambda_i) = - \prod_{j=1}^M e_2(\lambda_i -\lambda_j)\ \prod_{j=1}^2 g_{2}(\lambda_i - \bar \lambda_j). \label{set1}
\ee
There is a second set of BAE describing the breather with rapidity $\bar \lambda_1$
\be
e_y(\bar \lambda_1 -\Theta)\ e_1^N(\bar \lambda_1) = - \prod_{j=1}^M g_2(\bar \lambda_1 -\lambda_j)\ e_{2}(\bar \lambda_1 - \bar \lambda_2).
\label{set2}
\ee
As already mentioned the second set is necessary for our purposes here, given that it facilitates the computation of the energy and momentum of the breather as well as the formulation of the corresponding quantization condition.

From the first set of BAE (\ref{set1}) the following density regarding the negative parity strings arises,
\be
\sigma(\lambda) = \sigma_0(\lambda) + {1\over N} \Big ({\mathrm B}(\lambda-\Theta) + \sum_{j=1}^2 {\mathrm R}(\lambda -\tilde \lambda_j) \Big ),
\ee
where we define the Fourier transforms of ${\mathrm R},\ {\mathrm B}$ as
\be
\hat {\mathrm R}(\omega) = - {\cosh({\omega \over 2}) \over \cosh((\nu-1){\omega \over 2})}, ~~~~~\hat {\mathrm B}(\omega) = {\sinh({y\omega \over 2}) \over 2\sinh ({\omega \over 2})\cosh((\nu-1){\omega \over 2})}
\ee
The second set (\ref{set2}) leads to the density describing the breather
\be
\bar \sigma(\lambda) = \bar \sigma_0(\lambda) + \Big ( t_b(\lambda-\Theta) +\sum_{j=1}^2 r_b(\lambda - \bar \lambda_j) \Big ),
\ee
where the corresponding Fourier transforms read as
\be
\hat {\bar \sigma}_0(\omega) ={\cosh((\nu-2){\omega \over 2})\over \cosh((\nu-1){\omega \over 2})}, ~~~~~\hat r_b(\omega) = -{\cosh((\nu-3){\omega \over 2}) \over \cosh((\nu-1){\omega \over 2})},
~~~~~\hat t_b(\omega)= {\cosh((\nu-y-1){\omega \over 2}) \over \cosh((\nu-1){\omega \over 2})}. \label{four4}
\ee
Moreover, it may be shown as in the soliton case that
\be
\bar \sigma_0(\lambda) = \bar \varepsilon(\lambda), ~~~~~\bar \varepsilon(\lambda) =
{1 \over 2 \pi} {d \bar p(\lambda) \over d \lambda}
\ee
where $\bar \varepsilon$ and $\bar p$ are the energy and momentum of the lightest breather respectively.

Similarly, a quantization condition for the breather may be formulated
\be
\Big (e^{i\bar p(\lambda_1)N} \bar S(\bar \lambda_1, \bar \lambda_2, \Theta) -1 \Big  )| \bar \lambda_1, \bar \lambda_2 \rangle =0,
\ee
due to the factorization of the scattering process $\bar S(\bar \lambda_1, \bar \lambda_2, \Theta)= S_b^{(1,1)}(\bar \lambda_1, \bar \lambda_2)\ T_b^{(1)}(\bar \lambda_1, \Theta)$.
Comparison of the latter formula with $\bar \sigma$ leads to the expressions for the breather scattering amplitude as well as the corresponding breather transmission amplitude:
\be
S_b^{(1, 1)}(\lambda) = \exp \Big [ - \int_{-\infty}^{\infty}\ {d\omega \over \omega} e^{-i \omega \lambda} \hat r_b(\omega)\Big ],~~~~~~
T_b^{(1)}(\hat \lambda) = \exp \Big [ - \int_{-\infty}^{\infty}\ {d\omega \over \omega} e^{-i \omega \hat \lambda} \hat t_b(\omega)\Big ] \label{exp4}
\ee
$\lambda = \bar \lambda_1 -\bar \lambda_2,\ \hat \lambda = \bar \lambda_1 -\Theta$.

Bearing in mind (\ref{four4}), (\ref{exp4}) as well as the useful identity (\ref{use1}) we conclude that the breather scattering amplitude  is given by the following hyperbolic ratios,
\be
S_b^{(1, 1)}(\theta) = - {\sinh({\theta \over 2} - {i\pi \over 2 \gamma})\ \sinh({\theta\over 2} + {i\pi \over 2\gamma} +{i\pi \over 2} )
\over \sinh({\theta \over 2} +{i\pi \over 2 \gamma})\ \sinh({\theta\over 2} -{i\pi \over 2\gamma} -{i\pi \over 2} )},
\label{final}
\ee
where we define: $\theta =  {\pi \lambda \over \gamma}$. The scattering amplitude coincides of course with the familiar sine-Gordon breather quantity, and for the lightest breather, which we consider here this corresponds to the scattering amplitude of the scalar sinh-Gordon field \cite{zamo, corrigan, doikou-nepo-breather}.

Similarly, through (\ref{exp4}) the breather transmission amplitude may be derived as
\be
T_b^{(1)}(\hat \theta) =- {\sinh({\hat \theta  -\eta_1\over 2} - {i\pi \over 4})\ \sinh({\hat \theta -\eta_2\over 2} -{i\pi \over 4} )
\over \sinh({\hat \theta -\eta_1 \over 2} +{i\pi \over 4})\ \sinh({\hat \theta -\eta_2\over 2} +{i\pi \over 4})}
\label{breatherfinal}
\ee
$\hat \theta = {\pi \hat \lambda \over \gamma}$,
the constants $\eta_i$ have been defined in (\ref{ident2}), and it is clear that our expression (\ref{breatherfinal}) for the breather transmission amplitude coincides with the one identified in \cite{corrigan}.

The results on the scattering and transmission amplitudes may be generalized for higher $n$-breathers, which are represented by
$n$-positive parity strings with real centers $\bar \lambda_j$. More precisely, it is straightforward to see (see e.g. \cite{doikou-nepo-breather}) that the scattering between two generic $n_1,\ n_2$ breathers may be expressed as:
\be
S_b^{(n_1, n_2)}(\lambda) = \prod_{l_1 = 1}^{n_1}\ \prod_{l_2=1}^{n_2}\ S_b^{(1, 1)}\Big (\lambda + {i\over 2} (n_1 - n_2 -2 l_1 +2 l_2)\Big ),
\ee
while the transmission amplitude of an $n$-breather is
\be
T_b^{(n)}(\hat \lambda)= \prod_{l=1}^{n}\ T_b^{(1)}\Big (\hat \lambda + {i\over 2} (n +1 -2 l)\Big ).
\ee

The spin of the one $n$-breather state in the presence of the defect can be also evaluated, and turns out to be $S^z = 0$. The spin of the breather is as expected zero, recall that in the attractive regime as pointed out in the previous subsection the ``effective'' defect spin is zero. With this we conclude our analysis on the transmission matrices for the kinks (solitons) and breathers in the (an)isotropic Heisenberg models.

\section{Discussion}
We have studied  in the present article one-dimensional (an)isotropic Heisenberg chains in the presence of a single impurity. We have recalled the construction of the kink $S$-matrix and have produced the transmission amplitudes through the Bethe ansatz equations. We have also been able to derive the breather's transmission matrix in the attractive regime of the XXZ chain. Our findings in the attractive regime of the XXZ model coincide with earlier results obtained in the context of the sine-Gordon model \cite{corrigan}, implying that the picture is consistent. Our findings in the XXX case as well as in the repulsive regime of the XXZ model are novel as far as we can tell and involve finite representations of the $\mathfrak{sl}_2$ and $\mathfrak{U}_q(\mathfrak{sl}_2)$. As a further check, comparison with relevant results at the classical level should be made, especially with studies in the Landau-Lifshitz \cite{doikou-karaiskos-LL}, and sine-Gordon models \cite{avan-doikou-defect2}.

Some comments on future directions are in order here. We have based our formulation here on finite representations of $\mathfrak{sl}_2,\ \mathfrak{U}_q(\mathfrak{sl}_2)$. More precisely, we have restricted our analysis on defect matrices, that correspond to finite representations, thus the algebraic Bethe ansatz formulation may be applied. The chosen representations possess highest weight states, and therefore the algebraic Bethe ansatz variation can be successfully applied. However, infinite dimensional representations of $\mathfrak{sl}_2,\ \mathfrak{U}_q(\mathfrak{sl}_2)$ for the defect matrix, can be also considered. In this case local gauge (Darboux) transformations in the spirit described in \cite{FT-XYZ} should be employed in order to extract the associated spectrum and BAE.

Moreover, there exist several open issues to be resolved; it is an intriguing task
to determine the string configurations as well as the positions of the real centers of
the $n$-strings within the XXZ model, in order to extract all the eigenvalues of the
trigonometric transmission matrix from the BAE. A natural generalization would also be to
extend our analysis in the case of higher rank (deformed) algebras. Finally, depending on the values of the coupling constant it is possible to consider the formation of bound states between the particle-like excitations and the defect. This analysis may be achieved via the investigation of the poles appearing in the overall physical factor of the transmission matrix. All the above are significant issues, which hopefully will be addressed in the near future.

\appendix

\section{Eigenvalues \& eigenstates of the defect matrix}
In the present appendix we identify the eigenvalues and the corresponding eigenstates
of the defect matrix. Note that the spin operator shares essentially the same
set of eigenstates. We begin with writing the defect matrix as
\[
 \tilde L(\lambda) = \sum_{m=1}^{{\mathrm n}}\Big( \mathcal{A}_m\ e^{(2)}_{11} \otimes e^{({\mathrm n})}_{mm} + \mathcal{D}_m \, e^{(2)}_{22} \otimes e^{({\mathrm n})}_{mm} \Big ) + \sum_{m=1}^{{\mathrm n}-1} \Big( \mathcal{B}_m \, e^{(2)}_{12} \otimes e^{({\mathrm n})}_{m+1m} +
 \mathcal{C}_m \, e^{(2)}_{21} \otimes e^{({\mathrm n})}_{mm+1} \Big)\, ,
\]
$\tilde L(\lambda) \in \mbox{End}\Big ({\mathbb C}^2 \otimes {\mathbb C}^{\mathrm n}[\lambda] \Big )$, $\ {\mathrm n} = 2 S +1$. We consider the following ansatz for the eigenstates:
\ba
&& |\psi_0 \rangle = \hat e^{(2)}_1 \otimes \hat e^{({\mathrm n})}_1, ~~~~~|\psi_{\mathrm n} \rangle = \hat e^{(2)}_{\mathrm n} \otimes \hat e^{({\mathrm n})}_{\mathrm n}\cr
&& |\psi_k \rangle =  f_k \, \hat e^{(2)}_1 \otimes \hat e^{({\mathrm n})}_{k+1} + f_{k+1} \,\hat  e_2^{(2)} \otimes \hat e^{({\mathrm n})}_k \, , \qquad k =1, \cdots ,n-1 \, , \label{eigens}
\ea
where $\hat e^{(d)}_k$ is a $d$-dimensional column vector with zero elements everywhere, and the unit at the $k$-th position.

Acting with $\tilde L$ on the vector, according to the rule $e^{(d)}_{ab}\ \hat e^{(d)}_c = \d_{bc}\ \hat e^{(d)}_a $, we find
\be
\tilde L(\lambda)\  |\psi_0 \rangle =(\lambda + {i{\mathrm n} \over 2})\ |\psi_0 \rangle, ~~~~~~ \tilde L(\lambda)\ |\psi_{{\mathrm n}}\rangle = (\lambda + {i {\mathrm n} \over 2})\ |\psi_{{\mathrm n}} \rangle \, ,
\ee
and
\be
\tilde L(\lambda)\ |\psi_k \rangle = \ldots
= f_k \, \hat e^{(2)}_1 \otimes \hat e^{({\mathrm n})}_{k+1} (\mathcal{A}_{k+1} + \mathcal{B}_k \,y_k)
+ f_{k+1} \hat e^{(2)}_2 \otimes \hat e^{({\mathrm n})}_k (\mathcal{D}_k + \mathcal{C}_k y_k^{-1}) =  \epsilon_{{\mathrm n}, k}\ |\psi_k \rangle\, ,
\ee
where we have defined $y_k \equiv \frac{f_{k+1}}{f_k}$. In order for $|\psi_k\rangle$ to be an eigenvector
we require the following relation to hold
\[
 \mathcal{A}_{k+1} + \mathcal{B}_k y_k  = \mathcal{D}_k + \mathcal{C}_k y_k^{-1}\,,
\]
or equivalently,
\[
\mathcal{B}_k \, y_k^2 + y_k \, (\mathcal{A}_{k+1}-\mathcal{D}_k) -\mathcal{C}_k =0~.
\]
Solving this algebraic equation provides us directly with the eigenvalues of the defect operator, which would
have the generic expression
\[
\e_{{\mathrm n},k}^{(1,2)} = \mathcal{A}_{k+1} + \mathcal{B}_k \, y_k^{(1,2)}~.
\]
\paragraph{Rational case:}
Let us first define the following functions in the rational case
\be
{\cal A}_k = \lambda +i \a_k +{i\over 2}, ~~~~~{\cal B}_k = {\cal C}_k = i C_k, ~~~~~{\cal D}_k = \lambda -i\a_k +{i\over 2}\, .
\ee
Substituting the corresponding functions for the rational case leads to the
following values of $y_k$
\[
 y_k^{(1)} =  \sqrt{\frac{k}{{\mathrm n}-k}} = -\frac{1}{y_k^{(2)}}\, ,
\]
implying that the corresponding eigenvalues are
\[
 \e_{{\mathrm n},k}^{(1,2)} = \l \pm \frac{i {\mathrm n}}{2} \, .
\]
It is interesting that the eigenvalues are $k$-independent, thus the generic spin-$S$
defect matrix possesses only two distinct eigenvalues, each one being ${\mathrm n}$-fold
degenerate.

\paragraph{Trigonometric case:} The relevant functions in this case are:
\ba
&& \mathcal{A}_k = {1\over 2} (e^{\mu(\l +\frac{i}{2})} q^{\a_k} - e^{-\m(\l + \frac{i}{2})}q^{-\a_k}),
 \qquad \mathcal{B}_k = \mathcal{C}_k = {1\over 2}(q-q^{-1})\tilde C_k, \cr
 && \qquad\qquad\qquad\quad  \mathcal{D} = {1\over 2}( e^{\mu(\l +\frac{i}{2})} q^{-\a_k} - e^{-\m(\l + \frac{i}{2})}q^{\a_k}).
\ea
After some calculations, we find that the eigenvalues in the trigonometric case
are much more complicated, and given by the following expression
\be
\e_{{\mathrm n},k}^{(1,2)} = \cos[\tfrac{\m}{2}({\mathrm n}-2k)]\sinh (\l\m) \pm
{1\over 2}\Big[-1 -\cos[\m(2k-{\mathrm n})]+2\cos({\mathrm n} \m) + \cosh 2\l\m (-1 + \cos[\m({\mathrm n}-2k)])\Big]^{\frac{1}{2}}\, .
\ee

\paragraph{Spin eigenvalues:}
We may also compute the total spin $S_T^z$ associated to the defect matrix
\be
S_T^z = {\sigma^z \over 2} \otimes {\mathbb I}_{{\mathrm n}} + {\mathbb  I}_2 \otimes S^z,
\ee
where $S^z$ is given by the ${\mathrm n} \times {\mathrm n}$ matrix (\ref{rep1}) corresponding to the spin $S$ representation,
and ${\mathbb I}_m$ are the ${\mathrm n}$-dimensional unit matrices. As noted already, the
states found above  (\ref{eigens}) are also the  $S^z$ eigenstates. The spin
eigenvalue problem reads then as
\ba
&& S_T^z \ |\psi_0 \rangle = \Big (S +{1\over 2} \Big) |\psi_0 \rangle\, , \cr
&& S_T^z\ |\psi_k \rangle = \Big (S +{1\over 2} -k \Big) |\psi_k \rangle, ~~~~~~k\in \Big \{ 1, \ldots, {\mathrm n}-1 \Big \} \cr
&& S_T^z\ |\psi_{\mathrm n} \rangle = \Big (-S  - {1\over 2}\Big) |\psi_n \rangle\, .
\ea
We end up with $2{\mathrm n}$ eigenstates and eigenvalues. Each eigenvalue $S-k +{1\over 2}$, $\ k \in \Big \{ 1, \ldots, {\mathrm n}-1 \Big \}$ is 2-fold degenerate,
so that  in total there are ${\mathrm n} +1$ distinct eigenvalues, as expected from
the spin summation rules, which  take the familiar values:
$S^z_T \in \Big \{ -S -{1\over 2}, \ldots, S+{1\over 2}\Big \}$.

\section{Unitarity, crossing symmetry \& Casimir operators}

In this appendix we present the unitarity and crossing symmetry properties of
the transmission matrix, and confirm the findings of the present investigation by
exploiting these basic requirements together with the fact that the transmission
matrix satisfies the quadratic algebra (\ref{rttb}). The unitarity and crossing symmetry
are given by the following expressions respectively
\ba
&& {\mathbb T}_{12}(\lambda)\ {\mathbb T}_{12}(-\lambda) ={\mathbb I}, \cr
&& {\mathbb T}_{12}^{t_1}(-\lambda +i)\ M_1\ {\mathbb T}_{12}^{t_1}(\lambda +i)\ M_1 = {\mathbb I} \, ,
\ea
where $t_1$ denotes transposition on the first vector space. In both the isotropic and
anisotropic case in the principal gradation, which are considered here, $M = {\mathbb I}$.

\paragraph{Rational case:}
Let us first consider the XXX isotropic case. Recall the expression found for the transmission matrix associated to the spin $S$ representation:
\ba
{\mathbb T}(\lambda, S) &=& {T(\lambda,S) \over i\lambda + S + {1\over 2} }\ {\mathbb M}(\lambda,S) \cr
{\mathbb M}(\lambda,S) &=& \begin{pmatrix}
 i\lambda + S^z+ { 1\over 2} &  S^- \cr
 S^+ &  i\lambda - S^z+ { 1\over 2}
\end{pmatrix}\, ,
\ea
where $T(\lambda)$ is defined in (\ref{exp1}).

The requirement for unitarity leads to:
\be
T(\lambda, S)\ T(-\lambda, S) = {\mathbb I}\, , \label{u1}
\ee
whereas the crossing symmetry condition provides:
\be
T(\lambda +i,S)\ T(-\lambda +i,S)\ {(i \lambda + S + {1\over 2})\ (-i \lambda + S + {1\over 2}) \over (i \lambda + S - {1\over 2})\
(-i \lambda + S - {1\over 2})} = {\mathbb I}. \label{u2}
\ee
To obtain the relations above we have used the following
\ba
&&{\mathbb M}_{12}(\lambda, S)\ {\mathbb M}_{12}(-\lambda, S) = {\mathbb M}^{t_1}_{12}(\lambda+i, S)\
{\mathbb M}^{t_1}_{12}(-\lambda+i, S) =\cr &&(\lambda^2 + C) {\mathbb I} = (i \lambda +S +{1\over 2})(-i \lambda +S +{1\over 2}){\mathbb I}
\ea
$C$ is the Casimir operator of $\mathfrak{sl}_2$, and one may easily show for the spin $S$ representation that
\be
C = (S^z)^2 + {1\over 2} \Big \{S^-,\ S^+\Big \} + {1\over 4} = {(2S+1)^2 \over 4}.
\ee
The latter is valid even for any generic real $S$. A discussion on the spin $S$ (any real) representation of $\mathfrak{sl}_2$
expressed in terms of differential operators may be found in e.g. \cite{review} and references therein.
It is easily confirmed by inspection that $T$ defined in (\ref{exp1}) satisfies
relations (\ref{u1}), (\ref{u2}) emanating from the basic properties.

\paragraph{Trigonometric case:}
Similarly for the anisostropic XXZ chain recall the transmission matrix associated to
the spin $S$ representation of the deformed algebra
\ba
{\mathbb T}(\lambda, \mu, S) &=& {T(\lambda,S) \over \sin (\mu (i\lambda +S+{1\over 2} ))}\ \tilde {\mathbb M}(\lambda, \mu, S) \cr
\tilde {\mathbb M}_q(\lambda, \mu,S) &=& \begin{pmatrix}
 \sin (\mu(i\lambda + S^z+ { 1\over 2})) &  \sin(\mu)\ S^- \cr
 \sin(\mu)\ S^+ &  \sin (\mu (i\lambda -S^z+ { 1\over 2}))
\end{pmatrix}
\ea
$\mu = \pi \gamma$, $T(\lambda,S)$ defined in (\ref{expb1}), (\ref{TT2}).

As in the isotropic case the unitarity and crossing-unitarity conditions lead to:
\ba
&& T(\lambda, \mu, S)\ T(-\lambda,\mu, S) = {\mathbb I}, \cr
&& T(\lambda +i,\mu,S)\ T(-\lambda +i,\mu,S)\ {\sin (\mu (i \lambda + S + {1\over 2}))\
\sin (\mu (-i \lambda + S + {1\over 2})) \over \sin(\mu (i \lambda + S - {1\over 2}))\
\sin( \mu (-i \lambda + S - {1\over 2}))} = {\mathbb I}
\ea
Again confirmation of the findings regarding the anisotropic case is easily done by inspection.

The latter relation were obtained using the fact that:
\ba
&&\tilde {\mathbb M}_{12}(\lambda, \mu, S)\ \tilde {\mathbb M}_{12}(-\lambda, \mu, S) =\tilde {\mathbb M}^{t_1}_{12}(\lambda+i,\mu, S)\ \tilde {\mathbb M}^{t_1}_{12}(-\lambda+i, \mu, S)=\cr &&
\Big (-{1\over 2} \cos(2 \mu i\lambda)+ {1\over 4} C_q \Big ){\mathbb I} = \sin (\mu(i \lambda +S +{1\over 2}))\ \sin(\mu(-i \lambda +S +{1\over 2})){\mathbb I}
\ea
$C_q$ is the associated {\it $q$-deformed} Casimir operator of the $\mathfrak{U}_q(\mathfrak{sl}_2)$ algebra ($q = e^{i\mu}$),
and one may easily show for the spin $S$ representation that
\be
C_q = q q^{2S^z} + q^{-1} q^{-2S^z} + (q -q^{-1})^2\ S^-\ S^+ = q q^{-2S^z} + q^{-1} q^{2S^z}
+ (q -q^{-1})^2\ S^+\ S^-= 2 \cos (\mu(2S+1)).
\ee
The results for both rational and trigonometric cases are valid for any real values $S$ of the representation.
In this case the representation may be expressed in terms of differential operators (see e.g. \cite{review} and references therein).

\section{The XXZ model: generic values of the spin}

In the main text we have computed the transmission matrices given that the
spin parameter takes values in a restricted interval. We generalize here our results for
generic spin values. It is clear that these generalizations are due to the periodicity of
the involved trigonometric functions.

\paragraph{Repulsive regime:}
Let us introduce the Fourier transform of $a_y$ for generic values of $y=2S$:
\be
\hat a_y(\omega) = {\sinh \Big (((2m +1)\nu - y){\omega \over 2}\Big ) \over \sinh (\nu {\omega \over 2})}, ~~~~2\ m\ \nu\ <\ y\ <\ 2\ (m+1)\ \nu.
\ee
The effect of the latter generalization has the following effects in our computations in the repulsive regime of the XXZ model:
the new ``shifted'' spin becomes
\be
\tilde S = S- m -{1\over 2}
\ee
whereas the expression for the transmission matrix is again given by (\ref{TT}), where
\ba
T(\hat \lambda, \gamma, \tilde S)= &=&  \prod_{k=0}^{\infty}
{\Gamma(\hat z +\gamma \tilde S -m +{\gamma \over 2}+(2k+1)\gamma)\ \Gamma(\hat z -\gamma \tilde S +m- {\gamma \over 2}+(2k+1)\gamma+1)\over \Gamma(\hat z +\gamma \tilde S -m +{\gamma \over 2}+2k\gamma)\ \Gamma(\hat z -\gamma \tilde S+m-{\gamma \over 2} +2(k+1)\gamma+1)} \cr
&\times & {\Gamma(-\hat z +\gamma \tilde S-m +{\gamma \over 2}+2k\gamma)\ \Gamma(-\hat z - \gamma \tilde S+m-{\gamma \over 2} +2(k+1)\gamma+1)\over \Gamma(-\hat z + \gamma \tilde S-m+{\gamma \over 2}+(2k+1)\gamma)\ \Gamma(-\hat z - \gamma \tilde S+m-{\gamma \over 2} +(2k+1)\gamma+ 1)}, \label{expbb1} \nonumber
\ea
$\hat z = i \hat\lambda \gamma$.

\paragraph{Attractive regime:}
First we introduce the generalized Fourier transform for $b_y$
\be
\hat b_y(\omega) = - {\sinh((y - 2m\nu){\omega \over 2})\over \sinh (\nu {\omega \over 2})}, ~~~~ m\ \nu\ <\ y\ <\  (m+1)\ \nu.
\ee
Then the transmission matrix in the attractive regime is given by (\ref{TT2}), where
\ba
T(\hat \lambda, \gamma, \tilde S) &=& \prod_{k=0}^{\infty}
{\Gamma(\hat z-\xi+m(\gamma+1)+2(k+1)\gamma +{1\over 2})\ \Gamma(\hat z+\xi-m(\gamma+1)+2k\gamma +{1\over 2})
\over \Gamma(\hat z-\xi+m(\gamma+1)+(2k+1)\gamma +{1\over 2})\ \Gamma(\hat z+\xi-m(\gamma+1)+(2k+1)\gamma +{1\over 2})}\cr &\times&
{\Gamma(-\hat z-\xi+m(\gamma+1)+(2k+1)\gamma +{1\over 2})\ \Gamma(-\hat z+\xi-m(\gamma+1)+(2k+1)\gamma +{1\over 2})
\over \Gamma(-\hat z-\xi+m(\gamma+1)+2(k+1)\gamma +{1\over 2})\ \Gamma(-\hat z+\xi-m(\gamma+1)+2k\gamma +{1\over 2})} \nonumber
\ea
where
\be
\hat z = i\hat \lambda, ~~~~~~\xi = S + {\gamma \over 2}, ~~~~\tilde S =m.
\ee


\begin{thebibliography}{1}

\bibitem{delmusi}
G. Delfino, G. Mussardo and P. Simonetti, ``Statistical models with a line of defect,'', Phys. Lett. {\bf B328} (1994) 123, {\tt hep-th/9403049};\\
G. Delfino, G. Mussardo and P. Simonetti,  ``Scattering theory and correlation functions in statistical models with a line of defect,'' Nucl. Phys. {\bf B432} (1994) 518, {\tt hep-th/9409076}.

\bibitem{konle}
R. Konik and A. LeClair, ``Purely Transmitting Defect Field Theories", Nucl. Phys {\bf B538} (1999) 587; {\tt hep-th/9703085}.

\bibitem{BCZ1}
P. Bowcock, E. Corrigan and C. Zambon, ``Some aspects of jump-defects in the quantum sine-Gordon model'', JHEP {\bf 08} (2005) 023, {\tt hep-th/0506169}.

\bibitem{corrigan}
E. Corrigan and C. Zambon, ``A transmission matrix for a fused pair of integrable defects in the sine-Gordon model'', J. Phys.  {\bf A43} (2010) 345201, {\tt arXiv:1006.0939 [hep-th]}.

\bibitem{bajnok}
Z. Bajnok, A. George, ``From Defects to Boundaries'', Int. J. Mod. Phys. A21 (2006) 1063, {\tt hep-th/0404199};\\
Z. Bajnok, ``Equivalences between spin models induced by defects'', J. Stat. Mech. 0606 (2006) P06010, {\tt hep-th/0601107};\\
Z. Bajnok, Zs. Simon, ``Solving topological defects via fusion'', Nucl. Phys. B802 (2007) 307, {\tt arXiv:0712.4292 [hep-th]}.

\bibitem{weston}
R. Weston, ``An Algebraic Setting for Defects in the XXZ and Sine-Gordon Models,'' {\tt arXiv:1006.1555 [math-ph]}.

\bibitem{annecydef1}
M. Mintchev, E. Ragoucy and P. Sorba, ``Scattering in the Presence of a Reflecting and Transmitting Impurity'', Phys. Lett. {\bf B547} (2002) 313, {\tt hep-th/0209052};\\
M. Mintchev, E. Ragoucy and P. Sorba, ``Reflection-Transmission Algebras'', J. Phys. {\bf A36} (2003) 10407, {\tt hep-th/0303187}.

\bibitem{annecydef2}
V. Caudrelier, M. Mintchev and E. Ragoucy, ``The quantum non-linear Schrodinger model with point-like defect'', J. Phys. {\bf A37} (2004) L367, {\tt hep-th/0404144}.

\bibitem{cozanls}
E. Corrigan and C. Zambon, ``Jump-defects in the nonlinear Schrodinger model and other non-relativistic field theories'', Nonlinearity {\bf 19} (2006) 1447, {\tt nlin/0512038}.

\bibitem{BCZ2}
P. Bowcock, E. Corrigan and C. Zambon, ``Affine Toda field theories with defects'', JHEP {\bf 01} (2004) 056, {\tt hep-th/0401020};\\
E. Corrigan and C. Zambon, ``Comments on defects in the $a_r$ Toda field theories'', J. Phys. {\bf A 42} (2009) 304008, {\tt arXiv:0902.1307 [hep-th]}.

\bibitem{BCZ3}
E. Corrigan and C. Zambon, ``On purely transmitting defects in affine Toda field theory'', JHEP {\bf 07} (2007) 001, {\tt arXiv:0705.1066 [hep-th]};\\
E. Corrigan and C. Zambon, ``A new class of integrable defects'', J. Phys. {\bf A 42} (2009) 475203;{\tt arXiv:0908.3126 [hep-th]}.


\bibitem{caudr}
V. Caudrelier, ``On a systematic approach to defects in classical integrable field theories'', IJGMMP {\bf 5} No. 7 (2008) 1085, {\tt arXiv:0704.2326 [math-ph]}.

\bibitem{haku}
I. Habibullin and A. Kundu, ``Quantum and classical integrable sine-Gordon model with defect'', Nucl. Phys. {\bf B795} (2008) 549, {\tt arXiv:0709.4611 [hep-th]}.

\bibitem{nemes}
F. Nemes, ``Semiclassical analysis of defect sine-Gordon theory'', Int. J. Mod. Phys. {\bf A 25} (2010) 4493; {\tt arXiv:0909.3268 [hep-th]}.

\bibitem{doikou-defect}
A. Doikou, ``Defects in the discrete non-linear Schrodinger model'', Nucl. Phys.  {\bf B854} (2012) 153, {\tt arXiv:1106.1602, [hep-th]}.

\bibitem{avan-doikou-defect1}
J. Avan and A. Doikou, ``Liouville integrable defects: the non-linear Schrodinger paradigm,'', JHEP {\bf 01} (2012) 040, {\tt arXiv:1110.4728 [hep-th]}.

\bibitem{avan-doikou-defect2}
J. Avan and A. Doikou, ``The sine-Gordon model with integrable defects revisited'', {\tt arXiv:1205.1661 [hep-th]}.

\bibitem{agui}
A.R. Aguirre, T.R. Araujo, J.F. Gomes and A.H. Zimerman, ``Type-II B\"acklund Transformations via Gauge Transformations,''  JHEP {\bf 12} (2011) 056,
{\tt arXiv:1110.1589 [hep-th]};\\
A.R. Aguirre, ``Inverse scattering approach for massive Thirring models with integrable type-II defects,'' J. Phys. A: Math. Theor. {\bf 45} (2012) 205205, {\tt arXiv:1111.5249 [math-ph]}.

\bibitem{doikou-karaiskos-LL}
A. Doikou and N. Karaiskos, ``Sigma models in the presence of dynamical point-like defects,'' Nucl. Phys. {\bf B867} [FS] (2013) 872, {\tt arXiv:1207.5503 [hep-th]}.

\bibitem{FT}
L.D. Faddeev and L.A. Takhtajan,
``Spectrum and scattering of excitations in the one-dimensional isotropic Heisenberg model,''
J.\ Sov.\ Math.\  {\bf 24} (1984) 241 [Zap.\ Nauchn.\ Semin.\  {\bf 109} (1981) 134];\\
L.D. Faddeev, Int. J. Mod. Phys. {\bf A10} (1995) 1845, {\tt hep-th/9404013}.

\bibitem{YBE}
P.P. Kulish and E.K. Sklyanin, Lecture Notes in Physics, Vol. 151, (Springer, 1982),pp. 61;\\
L.A. Takhtajan, Quamtum Groups, Introduction to Quantum Groups and Intergable Massive models
of Quantum Field Theory, eds, M.-L. Ge and B.-H. Zhao, Nankai Lectures on Mathematical
Physics, World Scientific, 1990, p.p. 69;\\
V.E. Korepin, N.M. Bogoliubov, and A.G. Izergin, Quantum Inverse Scattering
Method, Correlation Functions and Algebraic Bethe Ansatz (Cambridge University
Press, 1993).

\bibitem{AndreiJohannesson}
A. Tsvelik and P.B. Wiegmann, ``The Exact Results for Magnetic Alloys,'' Adv. Phys. {\bf 32}, 331
(1983) 17;\\
N. Andrei and H. Johannesson, ``Heisenberg chain with impurities (an integrable model),'' Phys. Lett. A {\bf 100} (1984) 108-112.

\bibitem{zvy} H. Frahm and A.A. Zvyagin, ``The open spin chain with impurity: an exact solution,'' J. Phys. Cond. Matt, 9 (1997) 9939;\\
A. Kluemper and A.A. Zvyagin, ``Disordered magnetic impurities in uniaxial critical quantum spin chains,'' J. Phys. Cond. Matt. 12 (2000) 8705;\\
A.A. Zvyagin, ``Finite Size effects in Correlated Electron Models: Exact Results,'' Imperial College Press (\& World Scientific) (2005).

\bibitem{review}
A. Doikou, S. Evangelisti, G. Feverati and N. Karaiskos, ``Introduction to Quantum Integrability'', Int. J. Mod. Phys. {\bf A25} (2012) 3307, {\tt arXiv:0912.3350 [math-ph]}.

\bibitem{doikou-nepo-sun}
A. Doikou and R.Nepomechie, ``Bulk and boundary S matrices for the SU(N) chain'', Nucl. Phys. {\bf B521} (1998) 547, {\tt hep-th/9803118}.

\bibitem{Andrei-Destri}
N. Andrei and C. Destri,  ``Dynamical Symmetry Breaking And Fractionization In A New Integrable Model,'' Nucl. Phys. B231 (1984) 445.

\bibitem{doikou-nepo-mezi2}
A. Doikou, L. Mezincescu and R. Nepomechie, ``Factorization of multiparticle scattering in the Heisenberg spin chain,'' Mod. Phys. Lett. {\bf A12} (1997) 2591, {\tt hep-th/9707155}.

\bibitem{Kulish-Resh}
P. P. Kulish and N. Yu. Reshetikhin, "Generalized Heisenberg ferromagnet and Gross--Neveu model," Zh. Eksp. Teor. Fiz., 80, No. I, 214-228 (1981)

\bibitem{doikou-nepo-mezi1}
A. Doikou, L. Mezincescu and R. Nepomechie, ``Simplified calculation of boundary S matrices,'' J. Phys. {\bf A30} (1997) L507, {\tt hep-th/9705187}.

\bibitem{kulres}
P.P. Kulish, N.Yu. Reshetikhin, J. Sov. Math. 23 (1983) 2435, Zap. Nauchn. Semin. 101 (1981) 101.

\bibitem{doikou-nepo-breather}
A. Doikou and R. Nepomechie, ``Direct calculation of breather S matrices,'' J. Phys. {\bf A32} (1999) 3663, {\tt hep-th/9903066}.

\bibitem{zamo}
A.B. Zamolodchikov and Al.B. Zamolodchikov,  ``Factorized S Matrices in Two-Dimensions as the Exact Solutions of Certain Relativistic Quantum Field Models,'' Ann. Phys. {\bf 120} (1979) 253.

\bibitem{doikou-nepo-critical}
A. Doikou and R. Nepomenchie,  ``Soliton S matrices for the critical $A^{(1)}_{N-1}$ chain,''
Phys. Lett. {\bf B462} (1999) 121, {\tt hep-th/9906069}.

\bibitem{FT-XYZ}
L.A. Takhtajan and L.D. Faddeev, ``The quantum inverse problem
method and the XYZ Heisenberg model,''Russian Math. Surveys 34:5, 13 (1979);\\
L.A. Takhtajan, ``The quantum inverse problem method and the XYZ Heisenberg model,'' Physica D3 (1981) 231.



\end{thebibliography}
\end{document}